# Tunneling anomalous Hall effect in a ferroelectric tunnel junction

M. Ye. Zhuravlev,[1,2,3] Artem Alexandrov,[2] L. L. Tao,[1] and Evgeny Y. Tsymbal [1,2]

[1] *Department of Physics and Astronomy, University of Nebraska, Lincoln, Nebraska 68588, USA*
[2] *Moscow Institute of Physics and Technology, Dolgoprudny, Moscow Region 141700, Russia*
[3] *St. Petersburg State University, St. Petersburg 190000, Russia*

We report on a theoretical study on the tunneling anomalous Hall effect (TAHE) in a ferroelectric tunnel junction (FTJ), resulting from spin-orbit coupling (SOC) in the ferroelectric barrier. For ferroelectric barriers with large SOC, such as orthorhombic $HfO_2$ and $BiInO_3$, we predict values of the tunneling anomalous Hall conductivity (TAHC) measurable experimentally. We demonstrate strong anisotropy in TAHC depending on the type of SOC. For the SOC with equal Rashba and Dresselhaus parameters, we predict the perfect anisotropy with zero TAHC for certain magnetization orientations. The TAHC changes sign with ferroelectric polarization reversal providing a new functionality of FTJs. Conversely, measuring the TAHC as a function of magnetization orientation offers an efficient way to quantify the type of SOC in the insulating barrier. Our results provide a new insight into the TAHE and open avenues for potential device applications.

Since its discovery more than a century ago,[1] the anomalous Hall effect (AHE) [2] has been attracting continued interest. Two distinct mechanisms of the anomalous Hall conductivity are commonly accepted: intrinsic and extrinsic. Both originate from broken time reversal symmetry and spin-orbit coupling (SOC), but the former is driven purely by the electronic band structure which gives rise to the spin-dependent transverse (anomalous) velocity[3] and the associated Berry curvature,[4] whereas the latter results from spin-dependent impurity scattering, such as the skew scattering [5] or the side jump scattering.[6]

Recently, the AHE was proposed in tunneling geometry and was coined the tunneling AHE (TAHE).[7-9] The TAHE can be observed in a tunnel junction, which consists of two metal electrodes, with one being ferromagnetic, separated by a thin barrier layer. The TAHE originates from the skew tunneling (in analogy to the skew scattering), where the spin-polarized carriers experience asymmetric chiral contributions to the tunneling transmission probability due to the SOC in the barrier or at the barrier/metal interface.[9]

The experimental demonstration of the TAHE is challenging due to the small SOC in the proposed conventional semiconductor barriers (~10 meV [10]). Recently, however, a number of ferroelectric materials have been predicted to exhibit a very large SOC (~$10^2$ – $10^3$ meV) resulting from a large polarization-induced potential gradient. [11-18] In addition to the sizable SOC favorable for the experimental demonstration of the TAHE, these materials have the advantage of the reversible ferroelectric polarization which can be switched by an applied electric field. Since ferroelectric materials are non-centrosymmetric, the spin-momentum coupling linear in wave vector $k$ is allowed by symmetry, giving rise to the linear Rashba and Dresselhaus SOC in the bulk of these compounds.[19] As the result, reversal of ferroelectric polarization changes sign of the SOC parameter and thus that of the TAHE, which enables a nonvolatile electric field control of the TAHE.[11,16,18] This property adds a new functionality to a ferroelectric tunnel junction (FTJ), which is known to exhibit a tunneling electroresistance (TER) effect – a sizable change in resistance of the FTJ with polarization reversal.[20-23]

In this work, we employ the quantum-mechanical transport theory to calculate the TAHE in a FTJ with a ferromagnetic electrode. In contrast to the previous work[9] considering the interfacial Rashba [24] and cubic bulk Dresselhaus [25] SOC, we focus on the linear bulk SOC, which is appropriate for the ferroelectric barriers. Based on these calculations, we analyze the tunneling anomalous Hall conductivity (TAHC) dependent on the type and magnitude of SOC, the magnetization orientation, and the exchange coupling. We discuss the feasibility to observe the TAHE in FTJs in real experimental conditions.

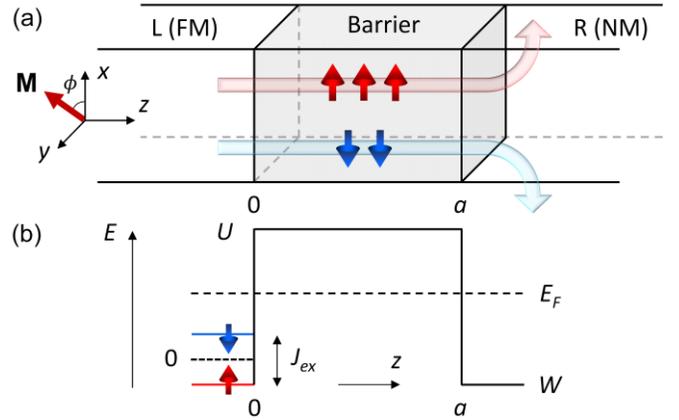

**FIG. 1.** (a) Schematic structure of a FTJ, which consists of semi-infinite left (L) ferromagnetic (FM) and right (R) nonmagnetic (NM) electrodes separated by a ferroelectric barrier of thickness $a$. FM magnetization **M** lies in the $x$-$y$ plane at angle $\phi$ with respect to the $x$ axis. Spin-dependent skew tunneling is schematically shown by the curved arrows indicating the two spin channels. (b) Potential profile across the junction. $E_F$ is the Fermi energy, $U$ is the barrier height, and $J_{ex}$ is the exchange splitting.

Fig. 1(a) shows a FTJ, which consists of a semi-infinite left (L) ferromagnetic (FM) electrode ($z < 0$) and a right (R) nonmagnetic (NM) electrode ($z > a$) separated by an insulating (ferroelectric) barrier of thickness $a$. The corresponding Hamiltonian in each region is given by

$$\begin{cases} H_L = -\dfrac{\hbar^2}{2m}\nabla^2 - \dfrac{J_{ex}}{2}(\sigma_x\cos\phi + \sigma_y\sin\phi), & z<0; \\ H_B = -\dfrac{\hbar^2}{2m}\nabla^2 + U + H_{SOC}, & 0<z<a; \quad (1) \\ H_R = -\dfrac{\hbar^2}{2m}\nabla^2 + W, & a>z. \end{cases}$$

Here $J_{ex}$ is the exchange splitting in the FM electrode, $\sigma_x$ and $\sigma_y$ are the Pauli matrices, $m$ is the electron effective mass, which is assumed to be constant in the whole junction. $U$ is the barrier height, $W$ is the potential in the NM electrode, and $\phi$ is magnetization angle with respect to the $x$ axis, as shown schematically in Fig. 1(b). The SOC in Eq. (1) is given by

$$H_{SOC} = \lambda_R(k_x\sigma_y - k_y\sigma_x) + \lambda_D(k_x\sigma_y + k_y\sigma_x), \quad (2)$$

which includes both the Rashba (the first term) and linear Dresselhaus (the second term) contributions.

The TAHC is determined by the spin-dependent scattering states in the NM electrode resulting from the incoming waves from both electrodes. The right propagating state of energy $E$ (normalized to the unit current density) incoming from the left FM electrode can be expressed as

$$\psi_L^\sigma = \sqrt{\dfrac{m}{\hbar k_z^\sigma}}\varphi_0 e^{ik_z^\sigma z}\chi_\phi^\sigma, \quad (3)$$

where $\varphi_0 = e^{i(k_x x + k_y y)}$, $\sigma = \uparrow,\downarrow$ is the spin index, $k_z^{\uparrow,\downarrow} = \sqrt{2m(E \mp J_{ex}/2)/\hbar^2 - k_\parallel^2}$ is the $z$-component of the wave vector, $\mathbf{k}_\parallel = (k_x, k_y)$ is the transverse wave vector, and $\chi_\phi^{\uparrow,\downarrow} = \dfrac{1}{\sqrt{2}}\begin{pmatrix} e^{-i\phi/2} \\ \pm e^{i\phi/2} \end{pmatrix}$ are the spinor eigenfunctions. Similarly, the left propagating state incoming from the right NM electrode is

$$\psi_R^\sigma = \sqrt{\dfrac{m}{\hbar q_z}}\varphi_0 e^{-iq_z z}\chi_\phi^\sigma, \quad (4)$$

where $q_z = \sqrt{2m(E-W)/\hbar^2 - k_\parallel^2}$. The scattering state in the right electrode due to the incoming state $\psi_L^\sigma$ is given by

$$\psi_{R\leftarrow L}^\sigma = \sqrt{\dfrac{m}{\hbar q_z}}\left(t_{RL}^{\sigma\sigma}e^{iq_z z}\chi_\phi^\sigma + t_{RL}^{\bar\sigma\sigma}e^{iq_z z}\chi_\phi^{\bar\sigma}\right), \quad (5)$$

where $\bar\sigma = -\sigma$ (i.e. $\bar\sigma = \downarrow$ if $\sigma = \uparrow$ and vice versa), and $t_{RL}^{\sigma\sigma}$ ($t_{RL}^{\bar\sigma\sigma}$) is the transmission amplitude between left and right electrodes without (with) spin flip. The scattering state in the right electrode due to the incoming state $\psi_R^\sigma$ is

$$\psi_{R\leftarrow R}^\sigma = \sqrt{\dfrac{m}{\hbar q_z}}\left[\left(e^{-iq_z z} + r_{RR}^{\sigma\sigma}e^{iq_z z}\right)\chi_\phi^\sigma + r_{RR}^{\bar\sigma\sigma}e^{iq_z z}\chi_\phi^{\bar\sigma}\right], \quad (6)$$

where $r_{RR}^{\sigma\sigma}$ ($r_{RR}^{\bar\sigma\sigma}$) is the reflection amplitude without (with) spin flip. Similarly, the scattering states in the left electrode due to the incoming states $\psi_L^\sigma$ and $\psi_R^\sigma$ are expressed as

$$\psi_{L\leftarrow L}^\sigma = \sqrt{\dfrac{m}{\hbar k_z^\sigma}}\left(e^{ik_z^\sigma z} + r_{LL}^{\sigma\sigma}e^{-ik_z^\sigma z}\right)\chi_\phi^\sigma + \sqrt{\dfrac{m}{\hbar k_z^{\bar\sigma}}}r_{LL}^{\bar\sigma\sigma}e^{-ik_z^{\bar\sigma}z}\chi_\phi^{\bar\sigma}, \quad (7)$$

$$\psi_{L\leftarrow R}^\sigma = \sqrt{\dfrac{m}{\hbar k_z^\sigma}}\left(t_{LR}^{\sigma\sigma}e^{-ik_z^\sigma z}\chi_\phi^\sigma + t_{LR}^{\bar\sigma\sigma}e^{-ik_z^{\bar\sigma}z}\chi_\phi^{\bar\sigma}\right), \quad (8)$$

respectively, were $r_{LL}^{\sigma\sigma}$ ($r_{LL}^{\bar\sigma\sigma}$) and $t_{LR}^{\sigma\sigma}$ ($t_{LR}^{\bar\sigma\sigma}$) are the respective reflection and transmission amplitudes. The scattering state in the barrier is given by

$$\psi_B = \varphi_0\begin{pmatrix} a_1^+ e^{Q_+ z} + a_2^+ e^{-Q_+ z} + a_1^- e^{Q_- z} + a_2^- e^{-Q_- z} \\ \gamma\left(a_1^+ e^{Q_+ z} + a_2^+ e^{-Q_+ z} - a_1^- e^{Q_- z} - a_2^- e^{-Q_- z}\right) \end{pmatrix}, \quad (9)$$

where $Q_\pm = \sqrt{2m(U-E\pm q)/\hbar^2 + k_\parallel^2}$, $\gamma = (i\alpha k_x + \beta k_y)/q$, and $q \equiv \sqrt{\alpha^2 k_x^2 + \beta^2 k_y^2}$.

The Hall current density $J_i^\sigma$ ($i=x,y$) resulting from $\psi_{R\leftarrow L}^\sigma$ can be expressed as

$$J_i^{L\sigma} = \dfrac{e}{(2\pi)^3\hbar}\int \text{Re}\left[(\psi_{R\leftarrow L}^\sigma)^* v_i \psi_{R\leftarrow L}^\sigma\right]f_L(1-f_R)dp, \quad (10)$$

where $dp \equiv d\mathbf{k}_\parallel dE$ and $v_i$ ($i=x,y$) is the velocity operator. $f_{L,R} = f(E-\mu_{L,R})$ and $\mu_{L,R}$ are, respectively, the Fermi function and the electrochemical potential of the left and right electrodes. Similarly, the Hall current resulting from $\psi_{R\leftarrow R}^\sigma$ is given by

$$J_i^{R\sigma} = \dfrac{e}{(2\pi)^3\hbar}\int \text{Re}\left[(\psi_{R\leftarrow R}^\sigma)^* v_i \psi_{R\leftarrow R}^\sigma\right]f_L(1-f_R)dp. \quad (11)$$

Substituting Eqs. (5) and (6) into Eqs. (10) and (11), we obtain

$$J_i^L = \dfrac{e}{(2\pi)^3\hbar}\sum_\sigma \int \dfrac{k_i}{q_z}\left(|t_{LR}^{\sigma\sigma}|^2 + |t_{LR}^{\bar\sigma\sigma}|^2\right)f_L(1-f_R)dp, \quad (12)$$

and $J_i^R = J_{i1}^R + J_{i2}^R$, where

$$J_{i1}^R = \dfrac{e}{(2\pi)^3\hbar}\sum_\sigma \int \dfrac{k_i}{q_z}\left[2\text{Re}(r_{RR}^{\sigma\sigma}e^{2iq_z z})\right]f_R(1-f_L)dp, \quad (13)$$

$$J_{i2}^R = \dfrac{e}{(2\pi)^3\hbar}\sum_\sigma \int \dfrac{k_i}{q_z}\left(1+|r_{RR}^{\sigma\sigma}|^2 + |r_{RR}^{\bar\sigma\sigma}|^2\right)f_R(1-f_L)dp. \quad (14)$$

It is notable that the current component $J_{i1}^R$ is $z$ dependent. This dependence originates from interference of the reflected waves incoming from the right electrode. Assuming that $\mu_{L,R} = E_F \pm eV/2$, where $E_F$ is the Fermi energy and $V$ is the bias voltage, at low temperature this component is zero for $V > 0$, but non-zero for $V < 0$. In the latter case, $f_R = 1$ and $f_L = 0$ in the energy window $[E_F - eV/2, E_F + eV/2]$ so that Eq. (13) is reduced to

$$G_{i1} = \frac{e^2}{(2\pi)^3 \hbar} \sum_\sigma \int \frac{k_i}{q_z} \left[ 2\text{Re}\left( r_{RR}^{\sigma\sigma} e^{2iq_z z} \right) \right]_{E=E_F} d\mathbf{k}_\parallel , \quad (15)$$

where conductance per unit area $G_{i1}$ is defined by $G_{i1} = J_{i1}^R / V$ and $V$ is assumed to be small. The integral of $G_{i1}$ over $z$ is zero, and thus $G_{i1}$ does not contribute to the total TAHC. However, the local variation of the TAHC is notable and discussed below.

The total Hall current is obtained by the sum of Eqs. (12) and (14) resulting in

$$J_i = \frac{e}{(2\pi)^3 \hbar} \sum_\sigma \int \frac{k_i}{q_z} \left( |t_{LR}^{\sigma\sigma}|^2 + |t_{LR}^{\bar\sigma\sigma}|^2 \right)(f_L - f_R) dp . \quad (16)$$

For small $V$ and low temperature, $f_L - f_R = \left(-\frac{\partial f}{\partial E}\right) eV = \delta(E - E_F) eV$ and the integration of Eq. (16) over $E$ leads to the TAHC per unit area $G_{iz} = J_i / V$ as follows

$$G_{iz} = \frac{e^2}{(2\pi)^3 \hbar} \sum_\sigma \int \frac{k_i}{q_z} \left( |t_{LR}^{\sigma\sigma}|^2 + |t_{LR}^{\bar\sigma\sigma}|^2 \right)_{E=E_F} d\mathbf{k}_\parallel , \quad (17)$$

which is in line with the previous result.[9] The respective transmission amplitudes can be obtained by matching the wave functions given by Eqs. (5)-(9) at the FTJ interfaces.

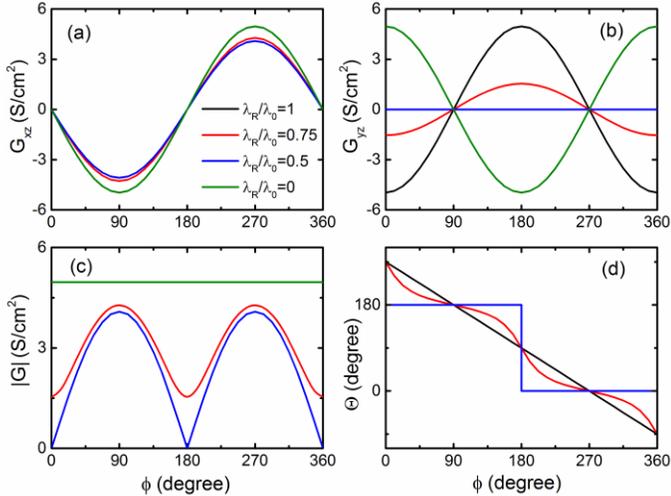

**FIG. 2.** Results of calculations of the TAHC as a function of magnetization angle $\phi$ for different SOC parameters, $\lambda_R$ and $\lambda_D$, such that $\lambda_R + \lambda_D = \lambda_0$, where $\lambda_0 = 1$ eV Å: (a) $G_{xz}$ component of TAHC, (b) $G_{xz}$ component of TAHC, (c) absolute value of TAHC $|G|$, and (d) angle $\Theta$ of the TAHE current with respect to the $x$-axis.

Next, we perform numerical calculations of the TAHC. In the calculations, we assume $a = 2$ nm, $E_F = 3$ eV, $U = 1$ eV, $W = -1$ eV, and $J_{ex} = 2$ eV as representative values. Fig. 2 shows the results for TAHC as a function of magnetization angle $\phi$ for different values of SOC parameters, $\lambda_R$ and $\lambda_D$, such that $\lambda_R + \lambda_D = \lambda_0$, where $\lambda_0 = 1$ eV Å. In agreement with the previous results,[9] we find that the Hall conductance $G_{xz}$ ($G_{yz}$) exhibits a sine(cosine)-type dependence on $\phi$. The TAHE originates from the imbalance of transmitted electrons with opposite transverse wave vectors, $\mathbf{k}_\parallel$ and $-\mathbf{k}_\parallel$, resulting from an effective spin- and $\mathbf{k}_\parallel$-dependent barrier height and the spin polarization of the FM electrode. The largest contribution to the TAHC occurs along directions where the spin polarization of the incoming electron is (anti)parallel to the polarization of the state at a given $\mathbf{k}_\parallel$. For example, electrons travelling along the $\mathbf{k}_\parallel = (0, k_y)$ direction and contributing to $G_{yz}$ tunnel through an effective spin-dependent barrier which height is determined by the SOC $(\lambda_D - \lambda_R) k_y \sigma_x$. In this case, the largest spin asymmetry in transmission is expected for electrons polarized along the $x$-direction, and hence the largest magnitude of $G_{yz}$ appears when the magnetization is (anti)parallel to the $y$-axis ($\phi = 0, 180°, 360°$ in Fig. 2 (b)). Rotating the magnetization changes the $x$-component of the spin such that $s_x \propto \cos\phi$, resulting in a cosine-type variation of $G_{yz}$. When $\lambda_D = \lambda_R$ and hence $(\lambda_D - \lambda_R) k_y \sigma_x = 0$, $G_{yz}$ vanishes (the green line in Fig. 2(b)). A similar interpretation is applied to the sine-type variation of $G_{xz}$ (Fig. 2 (a)). In this case, however, under conditions of $\lambda_R + \lambda_D = \lambda_0$ fixed, $G_{xz}$ is weakly dependent on $\lambda_R$ due to the spin-dependent tunneling barrier height at $\mathbf{k}_\parallel = (k_x, 0)$ being determined by $\lambda_0 k_x \sigma_y$ independent of $\lambda_R$.

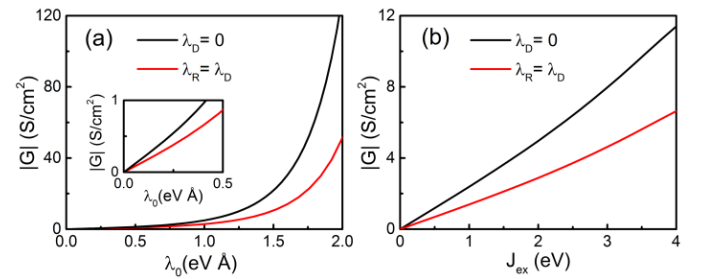

**FIG. 3.** TAHC $|G|$ as a function of (a) SOC parameter $\lambda_0$, where $\lambda_0 = \lambda_R + \lambda_D$, and (b) exchange coupling $J_{ex}$ for $\lambda_0 = 1$ eV Å. In case of $\lambda_D = \lambda_D$, the magnetization angle is fixed at $\phi = 45°$.

Fig. 2(c) shows the absolute value of TAHC $|G|$ as a function $\phi$. As expected, $|G|$ is $\phi$ independent for the pure Rashba or Dresselhaus SOC, while it varies notably with $\phi$ at intermediate values of SOC. Interestingly, $|G|$ becomes zero when the magnetization is parallel or antiparallel to the *x*-axis. This distinct $\phi$-dependent TAHC for different SOC points to the possibility of quantifying the SOC in a TAHE experiment. Fig. 2 (d) shows the angle $\Theta$ of the Hall current direction with respect to the *x*-axis. For the pure Rashba (or Dresselhaus) SOC, $\Theta$ is a linear function of $\phi$, while at intermediate SOC, it has a tendency to exhibit a step-like behavior consistent with the TAHC features discussed above. When the Rashba and Dresselhaus parameters are equal ($\lambda_R/\lambda_0 = 0.5$), $\Theta$ becomes a perfect step function of $\phi$, due to $G_{yz}$ being zero (Fig. 2(b)) while $G_{xz}$ exhibiting a sign change at $\phi = 180°$ (Fig. 2(a)).

The magnitude of the TAHC is largely controlled by the SOC and increases with increasing $\lambda_R$ or $\lambda_D$. As is evident from Fig. 3(a), the increase is linear at small values of SOC (insert in Fig. 3(a)), but at larger values of SOC ($\lambda_0 \geq 1$ eV Å) |G| increases exponentially with $\lambda_0$. The exchange coupling $J_{ex}$ determines the spin imbalance of the current carriers being responsible for TAHC. Therefore, as is seen from Fig. 3(b), |G| is zero in the absence of spin polarization, when $J_{ex} = 0$, but increases nearly linear with increasing $J_{ex}$.

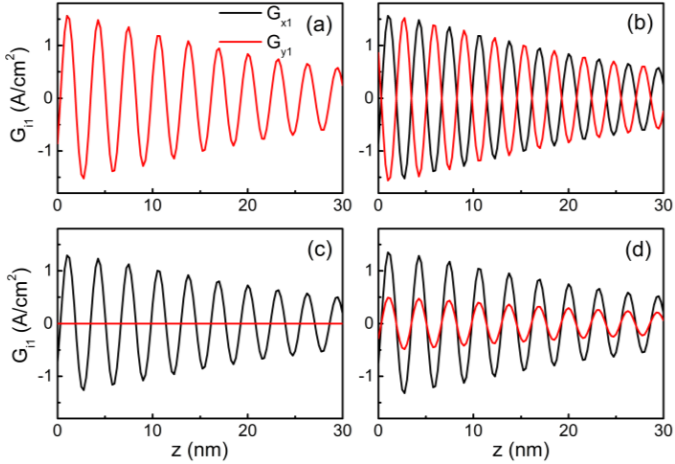

**FIG. 4.** The *z*-dependent components of the Hall conductance, $G_{x1}$ and $G_{y1}$, as a function of distance from the right interface for different SOC parameters, $\lambda_R$ and $\lambda_D$, such that $\lambda_R + \lambda_D = \lambda_0$ and $\lambda_0 = 1$ eV Å: (a) $\lambda_R/\lambda_0 = 1$, $\lambda_D = 0$, (b) $\lambda_R = 0$, $\lambda_D/\lambda_0 = 1$, (c) $\lambda_R/\lambda_0 = 0.5$, $\lambda_D/\lambda_0 = 0.5$, and (d) $\lambda_R/\lambda_0 = 0.75$, $\lambda_D/\lambda_0 = 0.25$. The magnetization orientation is fixed at $\phi = 45°$.

The presence of switchable ferroelectric polarization and large SOC in the tunnel barrier opens additional interesting possibilities for the TAHE. In ferroelectric materials, the spin texture is fully reversed in response to polarization switching.[11,16] This changes sign of the SOC parameters $\lambda_R$ and $\lambda_D$ in Eq. (2) resulting in reversal of the TAHC. The electrically switchable TAHC offers a new functionality of the FTJs which can be observed experimentally.

There are a number of ferroelectric oxide materials with a large SOC which can be employed for performing the TAHE experiment. For example, a large Rashba SOC $\lambda_R = 0.74$ eV Å was found in $BiAlO_3$.[13] A giant SOC with equal Rashba and Dresselhaus parameters $\lambda_R = \lambda_D = 0.96$ eV Å was predicted for $BiInO_3$.[18] If used in a FTJ, the latter would produce a perfect anisotropy in the TAHC with zero (non-zero) response for magnetization pointing along the *x*- (*y*-) direction. Another viable choice for a ferroelectric barrier is orthorhombic $HfO_2$,[26] where a large Dresselhaus SOC $\lambda_D = 0.58$ eV Å was predicted.[16] This material has been used as a barrier in FTJs showing a reversible polarization switching[27] as well as the TER effect.[28,29] One can estimate the Hall voltage $V_x$ for a FTJ with a ferroelectric $HfO_2$ barrier layer as follows:[9] $V_x \sim (G_{xz}/G_{el})V$, where $G_{el}$ is the conductance of the electrode. Assuming for simplicity a sample with equal tunneling and Hall contact areas $A \sim 10\times10$ μm$^2$, resistivity of the electrode $\rho \sim 10$ μΩ cm, and $V \sim 1$V, and taking into account the calculated $G_{xz} \sim 3$ S/cm$^2$, we find $V_x \sim 3$ nV, which is measurable experimentally.

Finally, we discuss the local variation of the TAHC resulting from the *z*-dependent Hall current contribution $G_{i1}$ given by Eq.(15). Fig. 4 shows the calculated $G_{i1}$ for different SOC parameters $\lambda_R$ and $\lambda_D$. It is seen that $G_{i1}$ exhibits an oscillatory behavior and a decay away from the interface. The oscillation period is determined by the Fermi wave vector $q_z$ in the right electrode. The slow decay $\propto z^{-1}$ results from the integration over $\mathbf{k}_\parallel$. Similar to the total TAHC, $G_{i1}$ reveals spatial anisotropy which is strongly dependent on the type of SOC. For the magnetization orientation $\phi = 45°$, we see that $G_{x1}$ and $G_{y1}$ oscillate in phase for the Rashba SOC (Fig. 4(a)), whereas they oscillate in antiphase for the Dresselhaus SOC (Fig. 4(b)). For equal Rashba and Dresselhaus SOC, the conductance is perfectly anisotropic with $G_{y1}$ being zero but $G_{x1}$ being finite (Fig. 4(c)). In a general case, both $G_{x1}$ and $G_{y1}$ are finite and oscillate over a large distance from the interface (Fig. 4(d)). We note, however, that detecting the oscillatory TAHC is challenging due to diffuse scattering in real experimental conditions.

In summary, we have studied the TAHE in FTJs based on the quantum-mechanical theory of spin-dependent electronic transport. For ferroelectric barriers with large SOC, such as orthorhombic $HfO_2$ and $BiInO_3$, we find TAHC values measurable experimentally. We predict anisotropy in the TAHC which depends on the type of SOC and becomes perfect for the SOC with equal Rashba and Dresselhaus parameters, where TAHC vanishes for a certain magnetization orientation. The TAHC changes sign with ferroelectric polarization reversal providing a new functionality of FTJs. We hope that our findings will stimulate experimental studies of the TAHE in FTJs.

This work was financially supported by the Moscow Institute of Physics and Technology (MIPT) and by the Russian Science Foundation (Grant No. 18-12-00434). The research at University of Nebraska was supported by the National Science Foundation (NSF) through Nebraska MRSEC (NSF Grant No. DMR-1420645).